\documentclass[prd,showpacs,letterpaper,twocolumn]{revtex4}%
\usepackage{graphicx}
\usepackage{bm}
\usepackage{epsf}
\usepackage{rotating}
\usepackage{epsfig,graphics,rotate,color}
\usepackage{wrapfig}
\usepackage{amssymb}
\usepackage{amsmath}
\usepackage{amsfonts}
\usepackage{array,hhline,dcolumn}%
\providecommand{\U}[1]{\protect\rule{.1in}{.1in}}
\begin{document}
\title{Using $L/E$ Oscillation Probability Distributions}

\author{
        A.~A. Aguilar-Arevalo$^{12}$, 
        B.~C.~Brown$^{6}$, L.~Bugel$^{11}$,
        G.~Cheng$^{5}$, E.~D.~Church$^{16}$, J.~M.~Conrad$^{11}$,
        R.~Dharmapalan$^{1}$, 
        Z.~Djurcic$^{2}$, D.~A.~Finley$^{6}$, R.~Ford$^{6}$,
        F.~G.~Garcia$^{6}$, G.~T.~Garvey$^{9}$, 
        J.~Grange$^{7}$,
        W.~Huelsnitz$^{9}$, C.~Ignarra$^{11}$, R.~Imlay$^{10}$,
        R.~A. ~Johnson$^{3}$, G.~Karagiorgi$^{5}$, T.~Katori$^{11}$,
        T.~Kobilarcik$^{6}$, 
        W.~C.~Louis$^{9}$, C.~Mariani$^{15}$, W.~Marsh$^{6}$,
        G.~B.~Mills$^{9}$,
        J.~Mirabal$^{9}$,
        C.~D.~Moore$^{6}$, J.~Mousseau$^{7}$, 
        P.~Nienaber$^{14}$, 
        B.~Osmanov$^{7}$, Z.~Pavlovic$^{9}$, D.~Perevalov$^{6}$,
        C.~C.~Polly$^{6}$, H.~Ray$^{7}$, B.~P.~Roe$^{13}$,
        A.~D.~Russell$^{6}$, 
        M.~H.~Shaevitz$^{5}$, 
        J.~Spitz$^{11}$, I.~Stancu$^{1}$, 
        R.~Tayloe$^{8}$, R.~G.~Van~de~Water$^{9}$, 
        D.~H.~White$^{9}$, D.~A.~Wickremasinghe$^{3}$, G.~P.~Zeller$^{6}$,
        E.~D.~Zimmerman$^{4}$ \\
\smallskip
(The MiniBooNE Collaboration)
\smallskip
}
\smallskip
\smallskip
\affiliation{
$^1$University of Alabama; Tuscaloosa, AL 35487 \\
$^2$Argonne National Laboratory; Argonne, IL 60439 \\
$^3$University of Cincinnati; Cincinnati, OH 45221\\
$^4$University of Colorado; Boulder, CO 80309 \\
$^5$Columbia University; New York, NY 10027 \\
$^6$Fermi National Accelerator Laboratory; Batavia, IL 60510 \\
$^7$University of Florida; Gainesville, FL 32611 \\
$^8$Indiana University; Bloomington, IN 47405 \\
$^9$Los Alamos National Laboratory; Los Alamos, NM 87545 \\
$^{10}$Louisiana State University; Baton Rouge, LA 70803 \\
$^{11}$Massachusetts Institute of Technology; Cambridge, MA 02139 \\
$^{12}$Instituto de Ciencias Nucleares, Universidad Nacional Aut\'onoma de M\'exico, D.F. 04510, M\'exico \\
$^{13}$University of Michigan; Ann Arbor, MI 48109 \\
$^{14}$Saint Mary's University of Minnesota; Winona, MN 55987 \\
$^{15}$Virginia Tech; Blacksburg, VA 24061 \\
$^{16}$Yale University; New Haven, CT 06520\\
}

\begin{abstract}

This paper explores the use of $L/E$ oscillation probability distributions to compare experimental measurements 
and to evaluate oscillation models.  In this case, $L$ is the distance of neutrino travel and $E$ is a measure of the interacting neutrino's energy. While comparisons using allowed and excluded regions for oscillation model parameters are likely the only rigorous method for these comparisons, the $L/E$ distributions are shown to give qualitative information on the agreement of an experiment's data with a simple two-neutrino oscillation model.  In more detail, this paper also outlines how the $L/E$ distributions can be best calculated and used for model comparisons.  
Specifically, the paper presents the $L/E$ data points for the final MiniBooNE data samples and, in the 
Appendix, explains and corrects the mistaken analysis published by the ICARUS collaboration. 

\end{abstract}

\maketitle

\section{Introduction}

In a simple model of oscillations involving two neutrinos,  the oscillation
probability,
\begin{equation}
P_{osc} = \sin^22\theta \sin^2(1.27\Delta m^2 L/E) ,
\label{posc}
\end{equation}
depends upon two experimental parameters: $L$, the distance traversed by the neutrino from
production to detection, and $E$, the neutrino energy. There are also two
fundamental parameters: $\theta$, the mixing angle, and $\Delta m^2$,
the difference of the squared neutrino masses.     The oscillation
probability for a model with three
mostly active neutrinos of negligible mass and one mostly sterile
neutrino of mass $\sim 1$ eV$^2$,  called a ``3+1 Model,'' is
approximated by this formula in the case of short baseline neutrino
experiments, where $L/E\sim 1$ m/MeV.     Recently,  3+1 models have been motivated
by anomalous signals observed in multiple short baseline experiments \cite{global_fits}, including
from the MiniBooNE experiment\cite{MBPRL}.  Therefore, there is motivation to
present data sets from multiple experiments on the same plot for cross-comparison.

\begin{figure}[tbp]
\vspace{-0.25in}
 \centerline{\includegraphics[width=6.9cm]{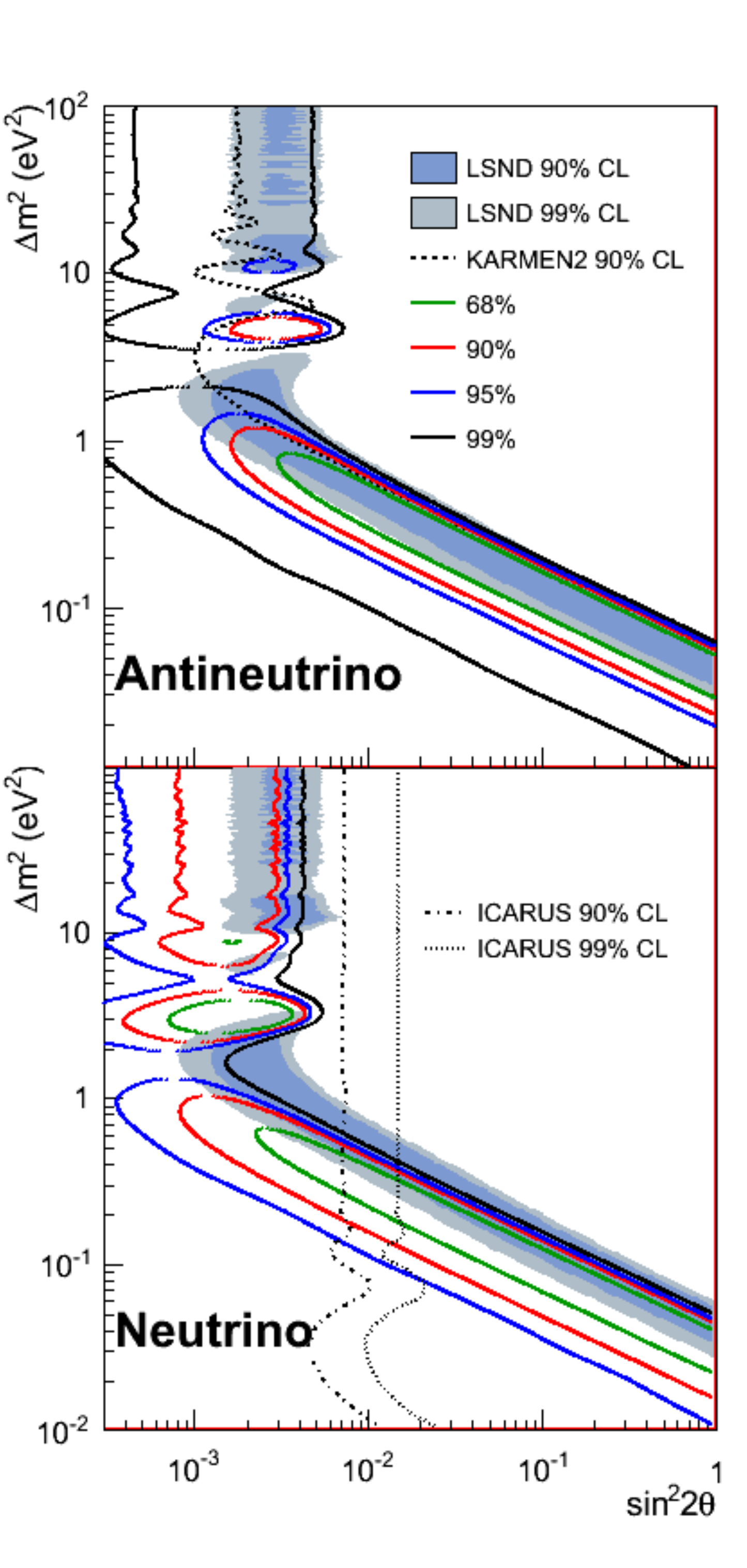}}
 \vspace{-0.3in}
\caption{MiniBooNE allowed regions in antineutrino mode (top) 
and neutrino mode (bottom) for events with
$E^{QE}_{\nu} > 200$ MeV within a two-neutrino oscillation model. 
Also shown are the ICARUS \cite{ICARUS} and KARMEN \cite{karmen} 
appearance limits for neutrinos and antineutrinos, respectively.
The shaded areas show the 90\% and 99\% C.L. LSND 
$\bar{\nu}_{\mu}\rightarrow\bar{\nu}_e$ allowed 
regions.}
\label{limitab}
\vspace{-0.1in}
\end{figure}

When comparing data
sets from experiments within such a model,  it is best to present the 
results 
in the $\Delta m^2$ versus $\sin^2 2\theta$ plane.  This method has
the advantage of incorporating all the information from the given experiments and putting the results on a common footing that can be rigorously compared.  
In particular, the distribution of ``true'' neutrino energies for any given ``reconstructed''
neutrino energy can be used to estimate the oscillation regions, and the
systematic uncertainties and correlations associated with neutrino flux, backgrounds, and
reconstruction at different energies can be correctly applied.
We strongly recommend that this method be used for comparing the results from oscillation experiments.  An example is shown in Fig.~\ref{limitab}, which is 
taken from Ref.~\cite{MBPRL} with some updates.  From this plot, one can easily and accurately compare the two-neutrino oscillation regions that are allowed and excluded by the various experiments.  

An alternative method overlays the distribution of
$P_{osc}$ versus $L/E$ from each experiment.  This method is
attractive because one can directly compare the experimental results
to Eq.~\ref{posc}, giving an intuitive sense of whether the results are
consistent with the expected oscillatory behavior.   However, this method is 
less accurate due to the introduction of additional
systematic errors associated with binning data in true $L/E$.   This makes
cross-comparison between experiments risky.    This paper
uses the MiniBooNE data set to illustrate these issues. 

In the $P_{osc}$ versus $L/E$ method,  one should bin the data in {\it
  true}, as opposed to {\it reconstructed},  $L/E$ and then calculate the measured oscillation probability for the events in
each bin.     This leads to multiple issues.    First, 
any given reconstructed event will be associated with a distribution of true $L/E$ values rather than a single,
definite value.  The potentially-very-wide distribution for true $L/E$
associated with an event is  due to many
sources.   With respect to the true energy,  uncertainties arise from
experimental resolutions on the reconstructed energies and angles of
outgoing particles.  Contribution also comes from unobserved particles in the event.
Substantial uncertainty on the distance can arise from 
the finite spatial extent of the neutrino source, especially in the
case of a decay-in-flight beams.  Second, when calculating 
the measured oscillation probability, $P_{osc}$, sizable
uncertainties may come from corrections for backgrounds and
reconstruction efficiencies.   
Lastly, one loses the
power of constraining systematic errors through use of correlation
error matrices, which are produced to be 
applicable to the full data set.

Other problems may also arise when presenting $P_{osc}$ versus $L/E$
plots.    For example, consider the analysis of
MiniBooNE data published by the ICARUS collaboration in
Ref. \cite{ICARUS}.   While ICARUS chose to present in $E/L$ rather
than $L/E$, the resulting oscillation probabilities clearly do not
agree with the already-published MiniBooNE analysis (Fig, 3 of Ref. \cite{arxiv1207.4809}).   
In the Appendix, we describe the mistake in the ICARUS analysis and
how it changes the derived MiniBooNE oscillation probabilities.   Note
that in the case of both of these analyses of the MiniBooNE data set,
the data are presented as a function of reconstructed $L/E$, not true
$L/E$.  Thus, beyond the mistake in calculated $P_{osc}$, 
cross comparing the results between the two experiments is not valid.

In this paper, we review the procedures and subtleties for correctly
determining the  oscillation probability  versus $L/E$.  
We provide tables of the MiniBooNE data points and errors.  Lastly,  we discuss the limitations of using
the oscillation probability versus the $L/E$ dependence when
comparing experiments and comparing to oscillation model predictions.

\section{The MiniBooNE Calculation of $P_{osc}$ vs. $L/E$}

The calculation of the oscillation probability as a function of $L/E$
makes use of the MiniBooNE $\nu_e$ and $\bar \nu_e$ data release
\cite{datarelease}, which was made available in 2012.   This provides a ``fully
oscillated'' Monte Carlo sample of electron-flavor events.  
This sample assumes 100\% $\nu_\mu \rightarrow \nu_e$
transmutation and corrects for the muon versus electron mass and cross section 
effects to give a correct electron neutrino event distribution.   For each fully
oscillated event, the release provides the reconstructed neutrino energy ($E_\nu^{QE}$), 
the true neutrino energy ($E_{true}$), the neutrino travel baseline, and an event weight.  This release also provides the MiniBooNE observed data and predicted background samples that are 
published in Ref.~\cite{MBPRL}.  These samples are binned in reconstructed charged-current quasielastic (CCQE) neutrino energy, $E_\nu^{QE}$, as described in Ref. \cite{arxiv1207.4809}, with the binning as described in the release.  For the $P_{osc}$ calculation, the ``fully oscillated''  Monte Carlo sample is also binned in these same reconstructed $E_\nu^{QE}$ bins.

The measured oscillation probability ($P_{osc}^{meas}$) 
in each reconstructed energy bin can then be calculated from the ratio of the data minus background excess to the ``fully oscillated'' prediction, as given by
\begin{equation}
(P_{osc}^{meas})_i = \frac{data_i-bkgnd_i}{fully\_oscillated_i} .
\label{pmeas}
\end{equation}
Using this formula yields the calculated MiniBooNE oscillation probabilities 
 given in Tables~\ref{Nuprobs} and \ref{Nubarprobs} along with the error that includes the statistical and systematic uncertainty associated with the excess.

\begin{table}
\centering
  \caption{\label{Nuprobs} The oscillation probability in reconstructed energy bins 
calculated for the MiniBooNE neutrino-mode data.
Also shown are the number of
excess events and the number of events expected for 100\% $\nu_\mu
\rightarrow \nu_e$ transmutation.}
  \begin{tabular}{ c | c | c | c }
    \hline
$E_\nu^{QE}$ (MeV)&Excess&100\% $\nu_\mu \rightarrow \nu_e$&$P_{osc}^{meas}$ \%\\
    \hline
200-300	&	$	52.7	\pm	22.8	$	&	4459	&	$	1.18	\pm	0.51	$	\\
300-375	&	$	53.1	\pm	13.9	$	&	5092	&	$	1.04	\pm	0.27	$	\\
375-475	&	$	36.9	\pm	15.3	$	&	9817	&	$	0.38	\pm	0.16	$	\\
475-550	&	$	13.8	\pm	10	$	&	8176	&	$	0.17	\pm	0.12	$	\\
550-675	&	$	-10.3	\pm	12.8	$	&	14600	&	$	-0.07	\pm	0.09	$	\\
675-800	&	$	2.9	\pm	10.5	$	&	13768	&	$	0.02	\pm	0.08	$	\\
800-950	&	$	-7.1	\pm	11.9	$	&	14169	&	$	-0.05	\pm	0.08	$	\\
950-1100	&	$	10.3	\pm	9.8	$	&	11103	&	$	0.09	\pm	0.09	$	\\
1100-1300	&	$	11.2	\pm	10.9	$	&	10613	&	$	0.11	\pm	0.1	$	\\
1300-1500	&	$	-2.5	\pm	8.7	$	&	6012	&	$	-0.04	\pm	0.14	$	\\
1500-3000	&	$	-0.9	\pm	14.5	$	&	6321	&	$	-0.01	\pm	0.23	$	\\
     \hline
  \end{tabular}
  \end{table}
  
  \begin{table}
\centering
  \caption{\label{Nubarprobs} The oscillation probability in reconstructed energy bins 
calculated for the MiniBooNE antineutrino-mode data.
Also shown are the number of
excess events and the number of events expected for 100\% $\bar\nu_\mu
\rightarrow \bar\nu_e$ transmutation.}
  \begin{tabular}{ c | c | c | c }
    \hline
$E_\nu^{QE}$ (MeV)&Excess&100\% $\bar\nu_\mu \rightarrow \bar\nu_e$&$P_{osc}^{meas}$ \%\\
    \hline
200-300	&	$	31.5	\pm	12.3	$	&	2336	&	$	1.35	\pm	0.53	$	\\
300-375	&	$	16.1	\pm	8.3	$	&	2562	&	$	0.63	\pm	0.32	$	\\
375-475	&	$	6.1	\pm	8.3	$	&	4228	&	$	0.15	\pm	0.2	$	\\
475-550	&	$	11.6	\pm	6	$	&	3563	&	$	0.32	\pm	0.17	$	\\
550-675	&	$	13.9	\pm	8.1	$	&	6022	&	$	0.23	\pm	0.13	$	\\
675-800	&	$	5	\pm	6.3	$	&	5612	&	$	0.09	\pm	0.11	$	\\
800-950	&	$	-0.5	\pm	6.9	$	&	5894	&	$	-0.01	\pm	0.12	$	\\
950-1100	&	$	-5.9	\pm	6.1	$	&	4707	&	$	-0.13	\pm	0.13	$	\\
1100-1300	&	$	-2.1	\pm	6.1	$	&	4383	&	$	-0.05	\pm	0.14	$	\\
1300-1500	&	$	5.4	\pm	5.2	$	&	2699	&	$	0.2	\pm	0.19	$	\\
1500-3000	&	$	0.6	\pm	9.4	$	&	2788	&	$	0.02	\pm	0.34	$	\\
     \hline
  \end{tabular}
  \end{table}

The oscillation probability in a given bin corresponds to neutrinos with a range of true neutrino energies and travel distances and so cannot be easily associated with one true $L/E$ value.  
One, therefore, needs to do some averaging over the events in the bin using a simulated event sample.  
However, the calculation of the average $L/E$ value corresponding to a given bin is not straightforward and, in fact, what kind of average to take is not clear.  One would like to calculate the $L/E$ value that corresponds to the above calculated oscillation probability for the distribution of events in the bin,
but this depends on the oscillation model assumed.  In the end, one needs to pick an approximation.  For the MiniBooNE results shown in Ref. \cite{arxiv1207.4809}, the choice, as described in the text, was to use $L/E = L_{avg}/E_\nu^{QE}$ with $L_{avg} = 525$ m \cite{OscProb}.   
While this may not be the best choice, it is straightforward and, as shown below, fairly accurate at higher
energies.

\begin{figure}[tbp]
\vspace{+0.1in}
\centerline{\includegraphics[angle=0, width=8.0cm]{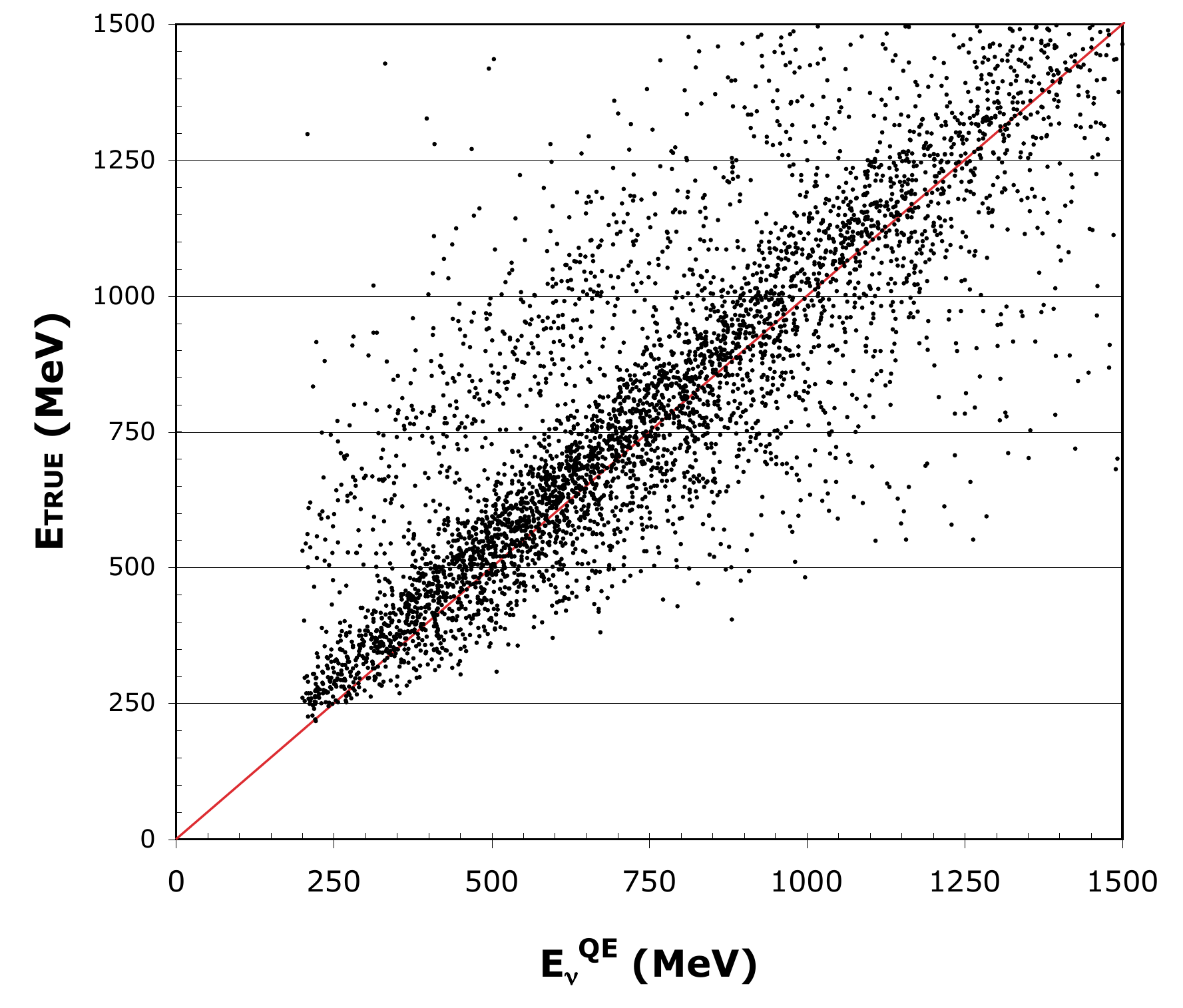}}
\vspace{-0.2in}
\caption{The energy distributions for neutrino-mode running for E$_{true}$ versus $E_\nu^{QE}$ for the ``fully oscillated'' sample.  The line indicates the case with perfect correlation.} 
\label{EtrueErecon}
\vspace{-0.2in}
\end{figure}

The main uncertainty in determining the best $L/E$ average in a reconstructed energy bin is the distribution of true energies for the events in the bin.  
For MiniBooNE, there can be a sizable shift in the mean due to the 
contamination of single pion events (CC1$\pi$) in the CCQE sample due to pion absorption in the nucleus.  This is modeled in the Monte Carlo event simulation, 
and uncertainties are included in all of the oscillation analyses.  Fig.~\ref{EtrueErecon} shows E$_{true}$ versus $E_\nu^{QE}$ for the ``fully oscillated'' event sample.  There is a clear shift to larger values of E$_{true}$ with respect to $E_\nu^{QE}$.
Also, there is a second band of events with E$_{true}$ higher than $E_\nu^{QE}$ by about 300 MeV, which corresponds to CC1$\pi$ events with an absorbed pion.  In principle, one could find the average E$_{true}$ for each of the $E_\nu^{QE}$ reconstructed energy bins for the ``fully oscillated'' event, but this would not be an appropriate average for an oscillation model with given $\Delta m^2$ and $\sin^2 2\theta$ values.   This change in the average energy is shown in 
Table~\ref{Eavgs}, where the average over the  ``fully oscillated'' sample for $E_\nu^{QE}$, E$_{true}$, and E$_{true}$ with several oscillation model weightings are presented.  The various averages show significant variations with sizable differences between the oscillation models, 
especially in the lower energy bins.   These variations in average energy then translate into changes in the average $L/E$ values for the bins as shown in 
Table~\ref{LovrE}, where an average over the ``fully oscillated'' sample is taken using the indicated energy parameter and true distance parameter to calculate $L/E$ for each event.  
Again, there are 10 - 20\% variations in the average $L/E$ values for different models out to reconstructed energy values of 800 MeV.  For the qualitative comparisons given in the next section, we will use the $L_{true}/E_{true}$  definition averaged over the ``fully oscillated'' sample for each reconstructed energy bin as the baseline $L/E$ variable.

\begin{table}[ht]
\centering
  \caption{\label{Eavgs}  The value of $E_\nu^{QE}$ and $E_{true}$ averaged over the ``fully oscillated'' sample with various weightings for the MiniBooNE neutrino-mode data.  
The Model 1 - 3 entries use $E_{true}$  but with weightings for three oscillations models with 
$\sin^22\theta/\Delta m^2 (eV^2) = 0.01/0.6, 0.004/1.0, $ and $0.002/2.0$.}
\begin{tabular}{r|rrrrr}
\hline
$E_\nu^{QE}$ (MeV)& $E_\nu^{QE}$ &  E$_{true}$ & Model 1 &  Model 2 & Model 3  \\
\hline
200 - 300	&	255	&	415	&	347	&	422	&	421	\\
300 - 375	&	341	&	465	&	408	&	442	&	532	\\
375 - 475	&	426	&	539	&	479	&	503	&	665	\\
475 - 550	&	513	&	607	&	548	&	567	&	691	\\
550 - 675	&	613	&	693	&	636	&	649	&	731	\\
675 - 800	&	737	&	793	&	734	&	746	&	808	\\
800 - 950	&	872	&	917	&	850	&	862	&	913	\\
950 - 1100	&	1021	&	1059	&	982	&	993	&	1040	\\
1100 - 1300	&	1193	&	1203	&	1115	&	1126	&	1173	\\
1300 - 1500	&	1388	&	1367	&	1259	&	1270	&	1321	\\
1500 - 3000	&	1761	&	1666	&	1444	&	1464	&	1549	\\
      \hline
\end{tabular}
 \end{table}

\begin{table}[ht]
\centering
  \caption{\label{LovrE} The $L/E$ values averaging over the ``fully oscillated'' sample with various definitions and weightings for the MiniBooNE neutrino-mode data.  
In all cases, the true neutrino travel distance for each event, $L_{true}$, is used in the averaging.  The Model 1 - 3 entries use $L/E_{true}$  but with weightings for three oscillations models with $\sin^22\theta/\Delta m^2 (eV^2) = 0.01/0.6, 0.004/1.0, $ and $0.002/2.0$.}
\begin{tabular}{c|ccccc}
\hline
 $E_\nu^{QE}$  & $L/E_\nu^{QE}$ & $L/E_{true}$ & Model 1 &    Model 2 &    Model 3 \\
\hline
200 - 300	&	2.088	&	1.480	&	1.646	&	1.389	&	1.494	\\
300 - 375	&	1.545	&	1.246	&	1.357	&	1.266	&	1.184	\\
375 - 475	&	1.234	&	1.056	&	1.152	&	1.099	&	0.872	\\
475 - 550	&	1.021	&	0.920	&	1.000	&	0.967	&	0.804	\\
550 - 675	&	0.854	&	0.794	&	0.857	&	0.837	&	0.742	\\
675 - 800	&	0.707	&	0.685	&	0.741	&	0.726	&	0.663	\\
800 - 950	&	0.597	&	0.590	&	0.640	&	0.629	&	0.586	\\
950 - 1100	&	0.507	&	0.507	&	0.554	&	0.546	&	0.513	\\
1100 - 1300	&	0.434	&	0.446	&	0.490	&	0.483	&	0.457	\\
1300 - 1500	&	0.373	&	0.392	&	0.438	&	0.432	&	0.409	\\
1500 - 3000	&	0.298	&	0.331	&	0.401	&	0.393	&	0.360	\\
\hline
\end{tabular}
  \end{table}
  
\section{Examples of using $P_{osc}$ vs. $L/E$}
  
With the caveats given above, one can try to use the calculated $P_{osc}$ versus $L/E$ values to compare the results from a given experiment to an oscillation model.  
Again, it should be stated that doing any oscillation fits using these types 
of data is fraught with inaccuracies associated with the variations of energy estimates, distance estimates, 
backgrounds, and the inclusion of correlated and uncorrelated systematic uncertainties.  On the other hand, comparisons using $P_{osc}$ versus $L/E$ variables can give qualitative information if the uncertainties can be minimized.   Comparing the data from several experiments using the $P_{osc}$ versus $L/E$ values can be even more difficult, as 
the relative shifts, inaccuracies, and systematic errors can be very different between the experiments.

\begin{figure}[tbp]
\vspace{+0.1in}
\centerline{\includegraphics[angle=0, width=8.5cm]{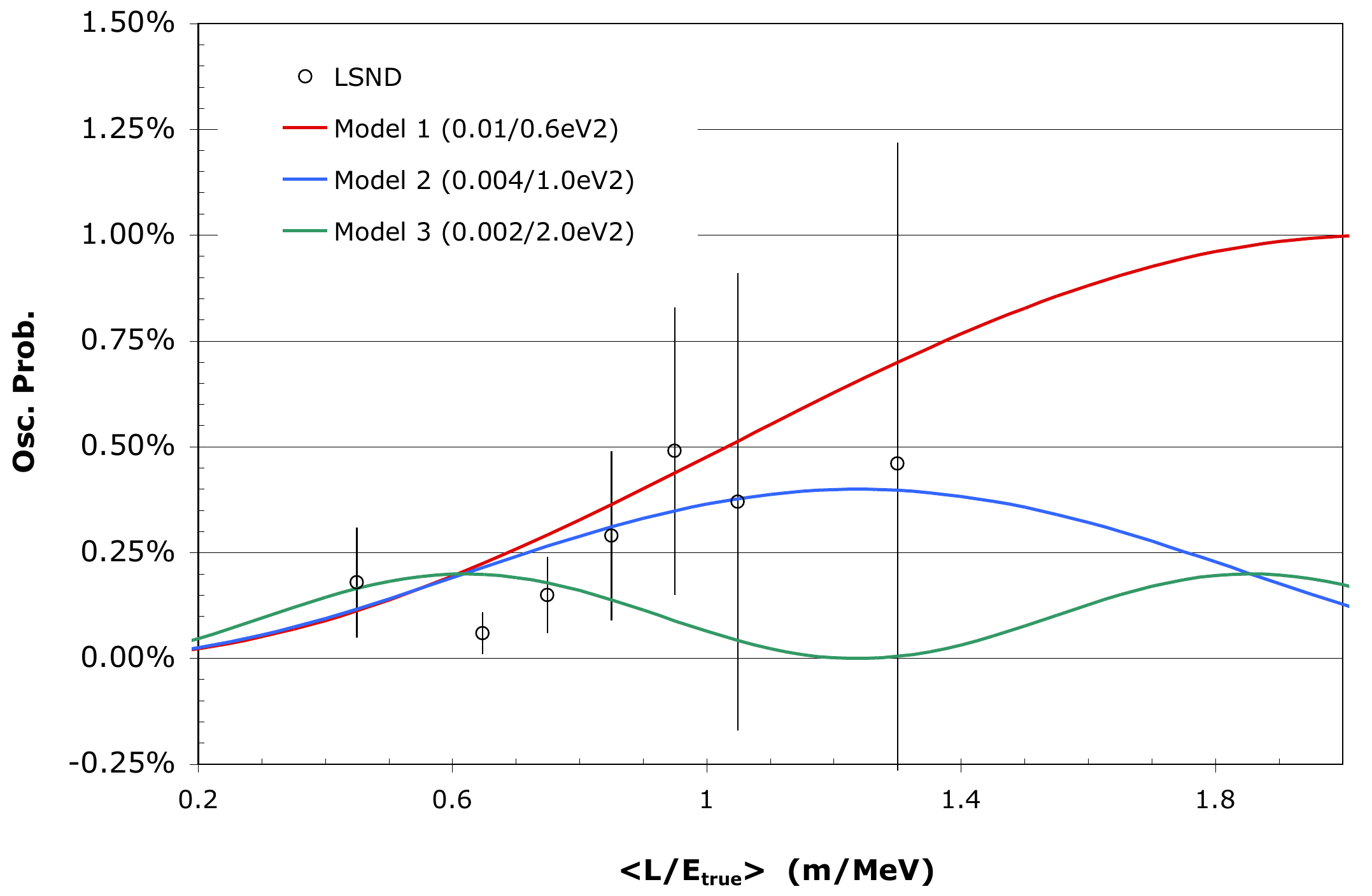}}
\vspace{-0.1in}
\caption{The LSND measured oscillation probability as a function of  the reconstructed $L/E$.   Three theoretical curves without any energy or flight path smearing are also shown for models with $\sin^22\theta/\Delta m^2 (eV^2) = 0.01/0.6, 0.004/1.0, $ and $0.002/2.0$. } 
\label{LSND_LovrE}
\vspace{-0.1in}
\end{figure}

Fig.~\ref{LSND_LovrE} shows the LSND $L/E$ data points from Ref.~\cite{arxiv1207.4809}.  The $L/E$ 
values for these data points were calculated using the LSND reconstructed energy.  As 
the LSND energy smearing is fairly Gaussian with a resolution at 52 MeV of 
$\sim 7\%$ and is well understood from the LSND muon decay calibration data, 
the measured $L/E$ from the reconstructed energy is accurate and does not produce sizable shifts.  For comparison, theoretical curves for three oscillation models in the LSND allowed regions from Fig.~\ref{limitab} are overlaid on the 
data points.  These theoretical curves have no energy or pathlength smearing applied, 
but, for LSND, these smearing effects are small.  The agreement between
the data points and theory curves is good; however, this 
agreement is tempered by the 
fact that the data points are correlated through the systematic uncertainties 
of the measurements.
  
\begin{figure}[tbp]
\vspace{+0.1in}
\centerline{\includegraphics[angle=0, width=8.5cm]{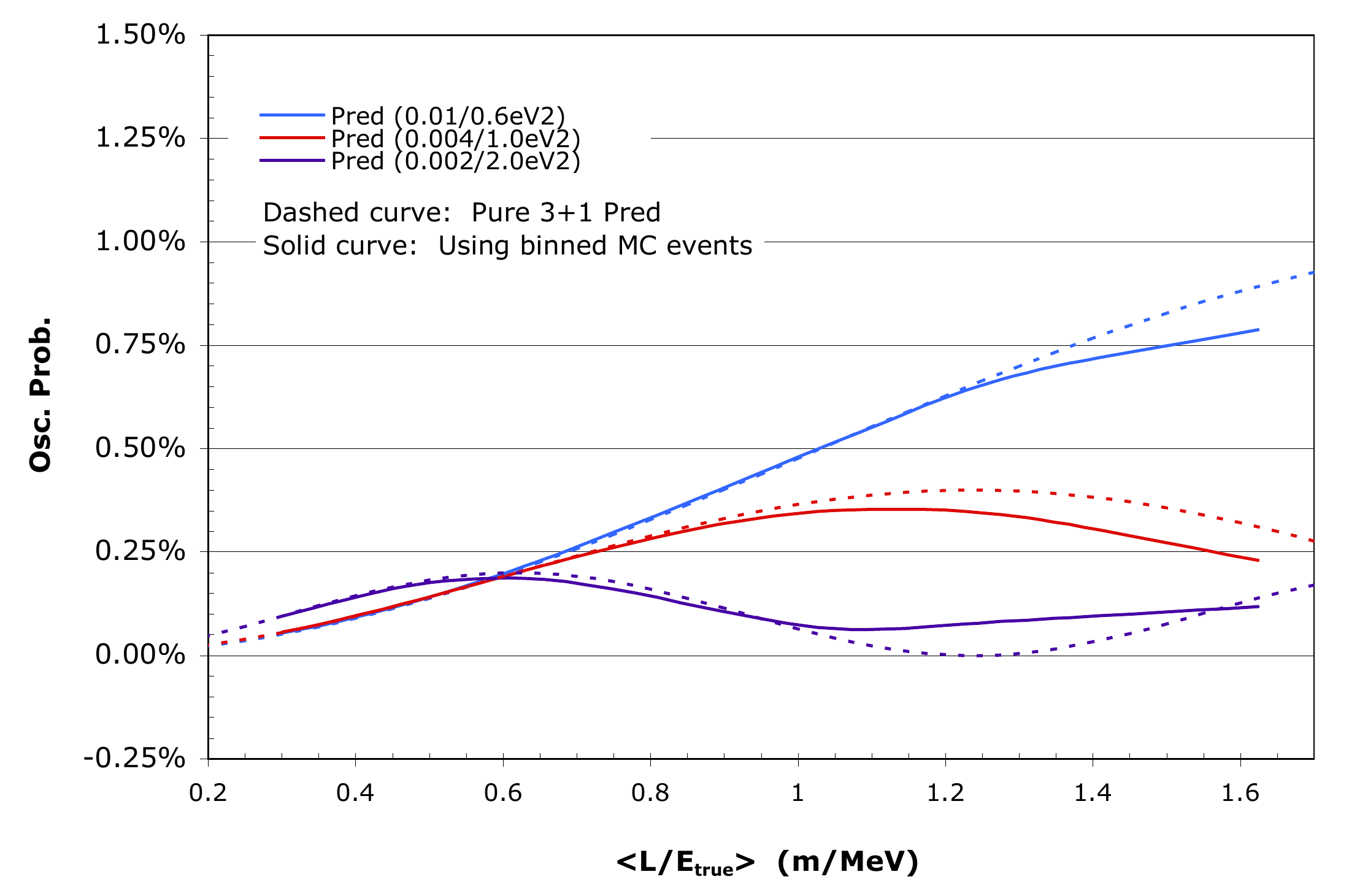}}
\vspace{-0.1in}
\caption{The predicted $P_{osc}$ versus baseline 
$L_{true}/E_{true}$ values (solid curves) with energy and flight path smearing 
for MiniBooNE neutrino-mode and antineutrino-mode running.  
The un-smeared theoretical predictions from Eq.~\ref{posc} are also shown (dashed curves).  The different curves correspond to oscillation models with $\sin^22\theta/\Delta m^2 (eV^2) = 0.01/0.6, 0.004/1.0, $ and $0.002/2.0$.} 
\label{Pred_curves}
\vspace{-0.1in}
\end{figure}

\begin{figure}[tbp]
\vspace{+0.1in}
\centerline{\includegraphics[angle=0, width=8.5cm]{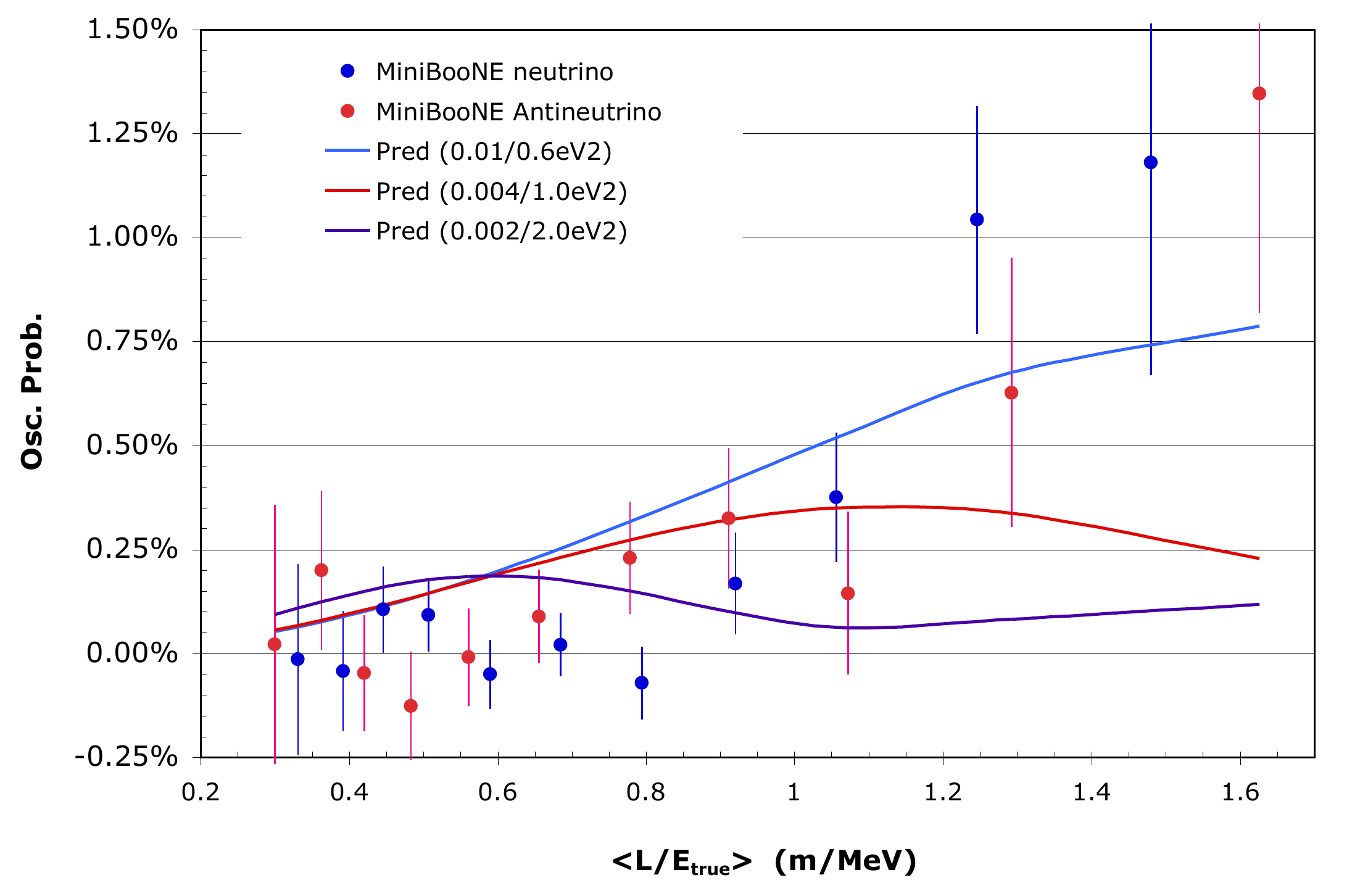}}
\vspace{-0.1in}
\caption{The  MiniBooNE $P_{osc}$ measurements as a function of the baseline $L_{true}/E_{true}$ variable for neutrino and antineutrino mode running.  
The curves are the predicted $P_{osc}$ versus baseline $L_{true}/E_{true}$
values with energy and flight path smearing.} 
\label{MB_with_Pred}
\vspace{-0.1in}
\end{figure}

For MiniBooNE, the energy and flight path smearing can be significant, 
and the predicted curves have deviations from the pure theoretical expectation.  This is shown in Fig.~\ref{Pred_curves}, where the dashed curves are the theoretical predictions from Eq.~\ref{posc}.
The solid curves are the prediction with smearing after integrating over the events in a reconstructed $E_\nu^{QE}$ bin using the ``fully oscillated'' sample weighted by the corresponding oscillation probability from Eq.~\ref{posc}.    As shown, the curves start to deviate at large $L/E$ values.  In order to compare the MiniBooNE $L/E$ points to oscillation models, one must use the smeared predictions of Fig.~\ref{Pred_curves}.  
Fig.~\ref{MB_with_Pred} shows the comparison of these predictions with the MiniBooNE data points for both neutrino-mode and antineutrino-mode running,
using the baseline $L_{true}/E_{true}$ variable.  The smeared prediction curves are almost identical for neutrino and antineutrino running, so only a single curve is displayed.  This plot should give a good representation of the MiniBooNE results with respect to these two-neutrino oscillation models.  
Only qualitative information can be taken from this comparison because of the strong correlations among the data points coming from systematic uncertainties.

\begin{figure}[tbp]
\vspace{+0.1in}
\centerline{\includegraphics[angle=0, width=8.5cm]{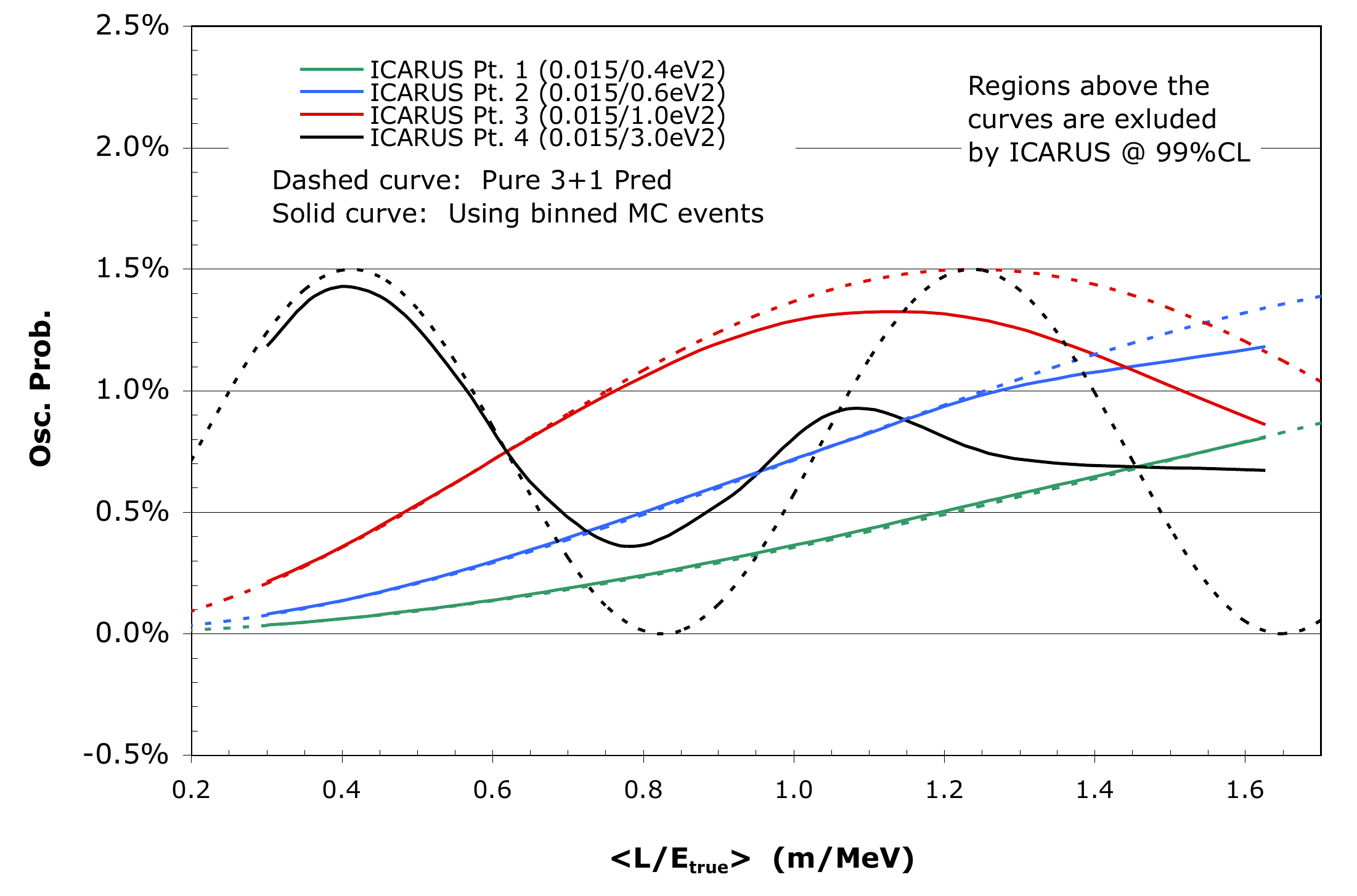}}
\vspace{-0.1in}
\caption{The predicted $P_{osc}$ versus baseline $L_{true}/E_{true}$ values
with energy and flight path smearing (solid curve) for MiniBooNE neutrino-mode 
and antineutrino-mode running. 
The un-smeared theoretical predictions from Eq.~\ref{posc} are also shown (dashed curves).  
The different curves correspond to oscillation models associated with the 
ICARUS 99\% C.L. limits with $\sin^22\theta = 0.015$ and $\Delta m^2  = 0.4, 0.6, 1.0$, and $3.0$ eV$^2$.} 
\label{Icarus_curves}
\vspace{-0.1in}
\end{figure}

\begin{figure}[tbp]
\vspace{+0.1in}
\centerline{\includegraphics[angle=0, width=8.5cm]{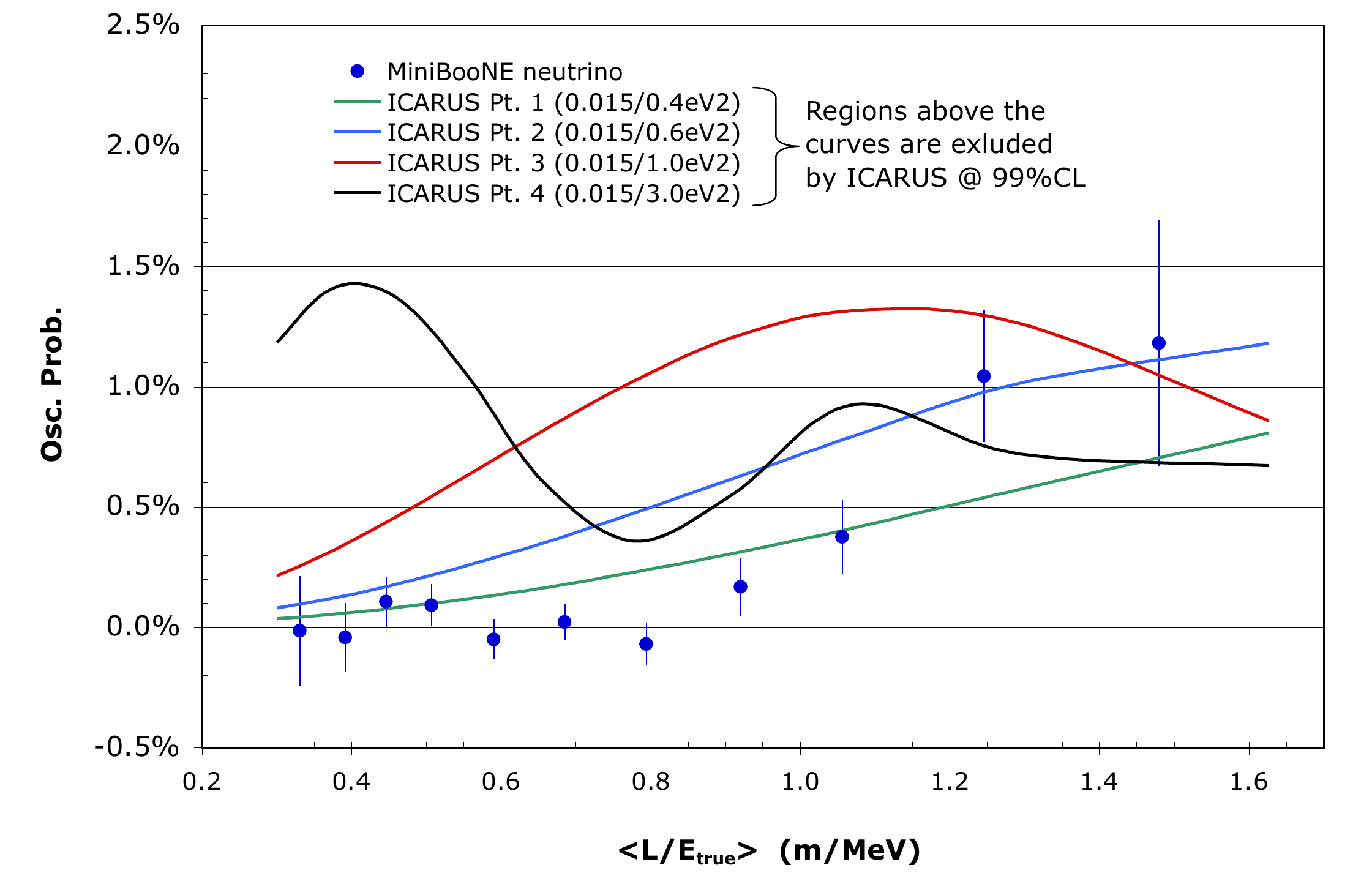}}
%\centerline{\includegraphics[angle=0, width=10.5cm]{Compare_MB_data_Icarus_limits_nu_only_revised.pdf}}
\vspace{-0.1in}
\caption{The  MiniBooNE $P_{osc}$ measurements (blue points) as a function of the baseline $L_{true}/E_{true}$ variable for neutrino-mode running. The curves are the 
ICARUS 99\% C.L. limits with $\sin^22\theta = 0.015$ and $\Delta m^2  = 0.4, 0.6, 1.0$, 
and $3.0$ eV$^2$ and with energy and flight path smearing.} 
\label{MB_with_Icarus_limits}
\vspace{-0.1in}
\end{figure}

The ICARUS collaboration searched for electron neutrino appearance in a muon neutrino beam and 
published a 99\% C.L. limit, as shown in Fig.~\ref{limitab}. 
Fig.~\ref{Icarus_curves} shows several example limit curves translated to the MiniBooNE setup both for the pure two-neutrino predictions and for the predictions including energy and flight path smearing.   Because the ICARUS experiment sets a limit, the region above the curves is excluded by their data at the 99\% C.L..  These limit curves can then be compared to the MiniBooNE neutrino data. 
As shown in Fig.~\ref{MB_with_Icarus_limits}, most of the MiniBooNE data 
points lie below the ICARUS curves and are, therefore, not excluded by their measurement.  
However, a quantitative statement would require including the correlated 
systematic uncertainties for the MiniBooNE data and would, in the end, just 
reproduce the comparison of the allowed and excluded regions shown in Fig.~\ref{limitab}.

\section{Conclusions}
With respect to searches for neutrino oscillations, the best method for comparing experiments and for comparing data to oscillation models is by determining the allowed or excluded regions with respect to an oscillation model.  
In this way, the details of the neutrino flux, event selection and reconstruction, and systematic uncertainties can be rigorously included.  For a 3+1 model, this will result in regions in the $\Delta m^2$ versus $\sin^22\theta$ plane.  Qualitative comparisons of an experiment's data to an oscillation model can also be approximated using the predicted oscillation probability versus $L/E$ value
from a model compared to a calculation of these quantities for the data.  This $L/E$ method is not exact due to smearing and other reconstruction effects, which are experiment dependent and preclude using this method to compare different experiments.  Further, it is difficult to use the $L/E$ analysis to do oscillation model fits due to the correlated uncertainties among the data points that would need to be taken into account.  
While the plot of the oscillation probability versus $L/E$ is a nice 
presentation of an experiment's data, it is not straightforward to use
for quantitative analysis.

\appendix

\section{Problems with the ICARUS $P_{osc}$ versus $L/E$ Analysis }

There are multiple issues associated with the ICARUS determination of the 
MiniBooNE oscillation probability.  ICARUS used the simple $L/E = L_{avg}/E_\nu^{QE}$ definition with $L_{avg} = 525$ m and the reconstructed energy $E_\nu^{QE}$ similar to Ref.~\cite{arxiv1207.4809}.  The major mistake in calculating the oscillation probability ($P_{osc}$) was to use the predicted excess shown in Fig.~2 of Ref.~\cite{MBPRL} for $\sin^22\theta=0.2$ and $\Delta m^2 = 0.1$ eV$^2$ along with the oscillation probability from Eq.~\ref{posc} evaluated at the reconstructed energy ($E_\nu^{QE}$) bin center.  ICARUS then divided the excess by this oscillation probability to try and obtain the number of ``fully oscillated'' events in the given bin.  To correctly calculate the ``fully oscillated'' events, one needs instead to divide by the oscillation probability calculated from the true energies of the events in a given reconstructed bin.  The ICARUS procedure using $E_\nu^{QE}$ significantly overestimates the oscillation probability in the bin for this test case, 
giving a smaller number of ``fully oscillated'' events ($N_{FullOsc}$).  
Calculating the $P_{osc}$ value for the bin from Eq.~\ref{pmeas} will give an erroneously high value,  where some data points are a factor of two off from the correct calculation.  
A comparison of the calculated $L/E$, $N_{FullOsc}$, and $P_{osc}$ values are shown in Table~\ref{IcarusProblem}.

Besides miscalculating the $P_{osc}$ values, the ICARUS publication also used incompatible $L/E$ values to compare the 
MiniBooNE data to their limit curves.  The ICARUS limit curves in Fig.~4 of 
their publication \cite{ICARUS} display theoretical curves for 
the predicted oscillation probability versus the true $L/E$ values.  These curves are compared to the MiniBooNE data plotted in terms of L/$E_\nu^{QE}$, which is significantly 
different, as shown in Table~\ref{IcarusProblem}.  As discussed above, the limit curves also need to be recalculated to take into account the true neutrino energies for the events in a given reconstructed energy bin.  If all of this was done properly, one would obtain the comparison as given in Fig.~\ref{MB_with_Icarus_limits}, which is very much different from the published ICARUS plot.

\begin{table}
\centering
  \caption{\label{IcarusProblem} A comparison of ``Correct'' and ``Incorrect ICARUS'' calculations of the oscillation probability in reconstructed energy bins for the MiniBooNE neutrino-mode data.
Also shown are the $L/E$ values and the number of events expected for 100\% $\nu_\mu
\rightarrow \nu_e$ transmutation ($N_{FullOsc}$).}
  \begin{tabular}{ c  c  c | c  c  c }
    \hline
 \multicolumn{3} {l|}{Correct Analysis} &  \multicolumn{3} {l}{Incorrect ICARUS Analysis}  \\
L/$E_{true}$&$N_{FullOsc}$&$P_{osc}^{meas}$ \%&L/$E_\nu^{QE}$&$N_{FullOsc}$&$P_{osc}^{meas}$ \% \\
    \hline
1.48	& 4459	&	$	1.18	\pm	0.51	$	&	2.09	&	2471	&	$	2.13	\pm	0.92	$	\\
1.25	& 5092	&	$	1.04	\pm	0.27	$	&	1.55	&	3501	&	$	1.52	\pm	0.40	$	\\
1.06	& 9817	&	$	0.38	\pm	0.16	$	&	1.23	&	7607	&	$	0.49	\pm	0.20	$	\\
0.92	& 8176	&	$	0.17	\pm	0.12	$	&	1.02	&	6972	&	$	0.20	\pm	0.14	$	\\
0.79	& 14600	&	$	-0.07	\pm	0.09	$	&	0.85	&	13106	&	$	-0.08	\pm	0.10	$	\\
0.69	& 13768	&	$	0.02	\pm	0.08	$	&	0.71	&	13418	&	$	0.02	\pm	0.08	$	\\
0.59	& 14169	&	$	-0.05	\pm	0.08	$	&	0.60	&	14171	&	$	-0.05	\pm	0.08	$	\\
0.51	& 11103	&	$	0.09	\pm	0.09	$	&	0.51	&	11147	&	$	0.09	\pm	0.09	$	\\
0.45	& 10613	&	$	0.11	\pm	0.1	$	&	0.43	&	11252	&	$	0.10	\pm	0.10	$	\\
0.39	& 6012	&	$	-0.04	\pm	0.14	$	&	0.37	&	7060	&	$	-0.04	\pm	0.12	$	\\
0.33	& 6321	&	$	-0.01	\pm	0.23	$	&	0.30	&	4326	&	$	-0.02	\pm	0.34	$	\\
     \hline
  \end{tabular}
  \end{table}
  
\begin{center}
{ \textbf{Acknowledgments}}
\end{center}
The authors thank the Department of Energy and the National Science
Foundation for support.


\begin{thebibliography}{99}                                                                   
\bibitem{global_fits}
 M.~Sorel, J.~M.~Conrad and M.~Shaevitz,
Phys.\ Rev.\ D {\bf 70}, 073004 (2004);
J.~M.~Conrad, C.~M.~Ignarra, G.~Karagiorgi, M.~H.~Shaevitz and J.~Spitz,
Adv.\ High Energy Phys.\  {\bf 2013}, 163897 (2013);
C.~Giunti and M.~Laveder, Phys.\ Lett.\ B 706, 200 (2011),
Phys.\ Rev.\ D84, 073008, (2011);
  J.~Kopp, P.~A.~N.~Machado, M.~Maltoni and T.~Schwetz,
 JHEP {\bf 1305}, 050 (2013).

\bibitem{MBPRL} 
A.~Aguilar-Arevalo {\em et~al.},
Phys.\ Rev.\ Lett.\ 110, 161801 (2013).

\bibitem{ICARUS} 
M.~Antonello {\em et~al.}, 
Eur.\ Phys.\ J.\ C73, 2599 (2013). 

\bibitem{karmen}
B.~Armbruster {\em et~al.},
Phys.\ Rev.\  D 65, 112001 (2002).

\bibitem{arxiv1207.4809} 
A.~Aguilar-Arevalo {\em et~al.},
arXiv:1207.4809.

\bibitem{datarelease} 
http://www-boone.fnal.gov/for\_physicists/data\_release/
nue\_nuebar\_2012/.

\bibitem{OscProb}
G.~B.~Mills,
``Oscillation Probability Versus L/E'',
MiniBooNE Technical Note 293 (2010).

\end{thebibliography}
\end{document}